\documentclass[aps,prx,twocolumn,superscriptaddress,showpacs,longbibliography,nofootinbib]{revtex4-1}
\usepackage{amsmath}
\usepackage{multirow}
\usepackage{graphicx}
\usepackage{amsfonts}
\usepackage{verbatim}
\usepackage{amssymb}
\usepackage{color}
\usepackage{amsmath} 
\usepackage{hyperref}
\hypersetup{colorlinks = true, urlcolor = blue, linkcolor = blue, citecolor = blue}
\usepackage[normalem]{ulem}

\newcommand{\su}{\uparrow} 
\newcommand{\sd}{\downarrow} 
\newcommand{\bpm}{\begin{pmatrix}}
\newcommand{\epm}{\end{pmatrix}}

\newcommand{\nn}{\nonumber \\} 

\newcommand{\dg}{^{\dagger}}

\newcommand{\half}{\frac{1}{2}}

\newcommand{\teff}{t_{\text{eff}}}
\newcommand{\Deff}{\Delta_{\text{eff}}}

\begin{document}

\title{Kitaev chain in an alternating quantum dot-Andreev bound state array}
\author{Sebastian Miles$^{*}$}
\affiliation{Qutech and Kavli Institute of Nanoscience, Delft University of Technology, Delft 2600 GA, The Netherlands}

\author{David van Driel}
\affiliation{Qutech and Kavli Institute of Nanoscience, Delft University of Technology, Delft 2600 GA, The Netherlands}

\author{Michael Wimmer}
\affiliation{Qutech and Kavli Institute of Nanoscience, Delft University of Technology, Delft 2600 GA, The Netherlands}

\author{Chun-Xiao Liu$^{*,\dagger}$}
\affiliation{Qutech and Kavli Institute of Nanoscience, Delft University of Technology, Delft 2600 GA, The Netherlands}

\date{\today}

\begin{abstract}
We propose to implement a Kitaev chain based on an array of alternating normal and superconductor hybrid quantum dots embedded in semiconductors.
In particular, the orbitals in the dot and the Andreev bound states in the hybrid are now on equal footing and both emerge as low-energy degrees of freedom in the Kitaev chain, with the couplings being induced by direct tunneling.
Due to the electron and hole components in the Andreev bound state, this coupling is simultaneously of the normal and Andreev types, with their ratio being tunable by varying one or several of the experimentally accessible physical parameters, e.g., strength and direction of the Zeeman field, as well as changing proximity effect on the normal quantum dots.
As such, it becomes feasible to realize a two-site Kitaev chain in a simple setup with only one normal quantum dot and one hybrid segment.
Interestingly, when scaling up the system to a three-site Kitaev chain, next-nearest-neighbor couplings emerge as a result of high-order tunneling, lifting the Majorana zero energy at the sweet spot.
This energy splitting is mitigated in a longer chain, approaching topological protection.
Our proposal has two immediate advantages: obtaining a larger energy gap from direct tunneling and creating a Kitaev chain using a reduced number of quantum dots and hybrid segments.
\end{abstract}

\maketitle

\def\thefootnote{*}\footnotetext{These authors contributed equally to this work}\def\thefootnote{\arabic{footnote}}
\def\thefootnote{$\dagger$}\footnotetext{{\color{blue}Corresponding author: chunxiaoliu62@gmail.com}}\def\thefootnote{\arabic{footnote}}

\section{Introduction}
The Kitaev chain is a toy model comprised of an array of spinless fermions with both normal and Andreev tunnelings between neighboring sites~\cite{Kitaev2001Unpaired}.
As a one-dimensional $p$-wave superconductor, the Kitaev chain in its topological phase will host a pair of Majorana zero modes localized at the endpoints of the chain~\cite{Alicea2012New, Leijnse2012Introduction, Beenakker2013Search, Stanescu2013Majorana, Jiang2013Non, Elliott2015Colloquium, Sato2016Majorana, Sato2017Topological, Aguado2017Majorana, Lutchyn2018Majorana, Zhang2019Next, Frolov2020Topological}.
These exotic quasiparticles are non-Abelian anyons, i.e., exchanging or braiding two Majoranas will transform between distinct ground-state wavefunctions in the degenerate manifold~\cite{Ivanov2001NonAbelian}.
Moreover, since two Majorana modes are spatially separated, quantum information encoded in such a Majorana pair will be more robust against local perturbation and decoherence.
With all these intriguing physical properties, Majorana zero modes are regarded as a promising candidate for implementing error-resilient topological quantum computing~\cite{Nayak2008Non-Abelian, DasSarma2015Majorana, Plugge2017Majorana, Karzig2017Scalable}.

In solid-state physics, one-dimensional topological superconductivity can be realized in several different types of hybrid materials, e.g., semiconductor-superconductor nanowires~\cite{Sau2010Generic, Lutchyn2010Majorana, Oreg2010Helical, Lutchyn2018Majorana, Mourik2012Signatures, Deng2016Majorana}, normal channels between planar Josephson junctions~\cite{Pientka2017Topological, Hell2017Two, Fornieri2019Evidence}, ferromagnetic atomic chains on top of a superconductor~\cite{Brydon2015Topological, Nadj-Perge14}.
Despite much experimental progress, a hybrid nanowire is inevitably subject to inhomogeneity and disorder, which can give rise to topologically trivial subgap states~\cite{Kells2012Near, Prada2012Transport, Liu2017Andreev, Moore2018Two, Reeg2018Zero, Vuik2019Reproducing, Pan2020Physical}, hindering an unambiguous detection of a topological superconductor.
Within this context, a very appealing solid-state platform for implementing a Kitaev chain is based on an array of semiconducting quantum dots~\cite{Sau2012Realizing}, which is much more immune to the effect of disorder owing to the large level spacing of dot orbitals relative to the disorder fluctuations.
In particular, under a sufficiently strong magnetic field, the spin-polarized dot orbitals serve as spinless fermions, coupling with neighboring ones through both, normal and Andreev couplings originating from elastic cotunneling (ECT), and crossed Andreev reflection (CAR) mediated by a superconductor.
Interestingly, even in a setup of only two quantum dots, a two-site Kitaev chain can be realized and host a pair of poor man's Majorana zero modes at a fine-tuned sweet spot~\cite{Leijnse2012Parity}.

Very recently, significant experimental progress has been made to transform the above-mentioned theoretical proposals and ideas into a physical realization.
In a minimal Kitaev chain device of double quantum dots, the conductance spectrocopies measured at the sweet spot are consistent with the signatures of Majorana zero modes~\cite{Dvir2023Realization}.
Here, the key physical insight is to mediate the effective couplings between dot orbitals using Andreev bound states (ABS) in a semiconductor-superconductor hybrid~\cite{Liu2022Tunable} instead of the continuum states of superconductivity~\cite{Sau2012Realizing, Leijnse2012Parity}.
Coupling through an ABS allows that the ratio of CAR and ECT amplitudes can be controlled by varying the chemical potential in the hybrid segment via electrostatic gating~\cite{Liu2022Tunable, Bordin2023Tunable, Wang2022Singlet, Wang2023Triplet}.
This effect was shown to be robust to Coulumb interactions in the dots as well as strong coupling~\cite{Tsintzis2022Creating}.
In spite of progress, current Kitaev chain devices are still suffering from several shortcomings which may limit its application in quantum technology in the future.
First, the excitation energy gap is relatively small ($\sim 25~\mu$eV), owing to the fact that CAR and ECT couplings, which are induced by second-order tunneling processes, scale with the tunneling amplitude as $\sim t^2_0/\Delta_0$, with $t_0$ the characteristic dot-hybrid tunneling strength and $\Delta_0$ the induced gap of ABS.
Second, when scaling up the system into an $N$-site Kitaev chain, one needs to have $N$ quantum dots and $N-1$ pieces of hybrid segments, which makes the device fabrication process increasingly challenging for a longer chain.

Alternatively to using normal quantum dots, Ref.~\onlinecite{Fulga2013Adaptive} proposed to use Andreev bound states in proximitized quantum dots directly as spinless fermions in a Kitaev chain. There, control over the proximity effect in each quantum dot was required, e.g. by using a quantum point contact to couple to the superconductor.

\begin{figure}[b]
    \centering
    \includegraphics[width=\linewidth]{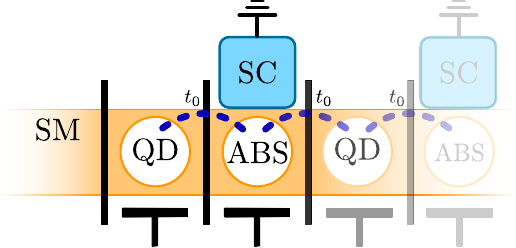}
    \caption{Schematic of a Kitaev chain device from alternating quantum dots and Andreev bound states created in semiconductor-superconductor hybrids.
    Quantum dots are formed by confinement potentials induced by tunnel gates (vertical black lines), while ABS emerge in the quantum dot hosted in the hybrid segment where the semiconductor is proximitized by superconductivity.
    The tunneling strength (purple dashed lines) between dot and ABS can be tuned by varying the voltage of the tunnel gates, and the chemical potentials of the dot and ABS can be adjusted by changing the plunger gate voltages (black T-elements). }
    \label{fig:device}
\end{figure}

In the current work, we propose a new method to create Kitaev chain combining the advantages of previous proposals. Our implementation is based on an array of alternating quantum dot and semiconductor-superconductor hybrid (see Fig.~\ref{fig:device}).
In particular, the orbitals in the quantum dots and the ABS in the hybrids are now on equal footing as the spinless fermions in the Kitaev chain, with the effective couplings being induced by direct tunneling.
Due to the electron and hole nature of the ABS, this coupling is simultaneously of the normal and Andreev type, with their ratio being tunable by varying one or several of the experimentally accessible physical parameters, such as strength and direction of the Zeeman field, as well as the changing the tunnel coupling between normal and hybrid quantum dots.
As such, it becomes possible to implement a two-site Kitaev chain in a simple setup with only one quantum dot and one hybrid segment, and in general an $N$-site Kitaev chain requires only $N$ pieces of basic elements of either dot or hybrid instead of $2N-1$ as proposed in Ref.~\cite{Sau2012Realizing}.
At the same time, our proposal does not require control of proximity effect in individual dots as in Ref.~\cite{Fulga2013Adaptive}, and can be realized in the same type of devices as previous experiments~\cite{Liu2022Tunable, Bordin2023Tunable, Wang2022Singlet, Wang2023Triplet}.
Moreover, the energy gap of the proposed Kitaev chain will be readily enhanced $\sim t_0$ owing to the direct tunneling between dot and hybrid.
Interestingly, when scaling up the system to a three-site Kitaev chain, next-nearest neighbor couplings emerge as a result of high-order tunneling, lifting the Majorana zero energy at the sweet spot. 
Nevertheless, this energy splitting is mitigated in a longer chain, giving robust zero mode within a larger parameter space, as topological protection is approached.

While our approach is based on alternating normal and hybrid quantum dots, a parallel work considers the case of two superconducting quantum dots showing that a phase difference alone can be used to tune to a sweet spot~\cite{Samuelson2023minimal}.

The remainder of the work is structured as follows: Section~\ref{sec:minimal_chain} focuses on the study of minimal Kitaev chain based on a single pair of quantum dot and ABS.
We introduce the model Hamiltonian in Sec.~\ref{sec:model_hamiltonian} and derive its low-energy effective theory in Sec.~\ref{sec:low_energy}.
In particular, in Sec.~\ref{sec:Zeeman_strength}-\ref{sec:proximity_dot} we show how one can systematically fine-tune the sweet spot using experimentally accessible physical parameters, e.g., strength and direction of the Zeeman field, as well as induced pairing gap on the quantum dots.
In Sec.~\ref{sec:scaling_up}, we consider scaling up of the dot-ABS chain, highlighting the emergence of next-nearest couplings and the effects on the Majorana properties at the sweet spot.
Section~\ref{sec:discussion} is devoted to discussions before we summarize our work in Sec.~\ref{sec:summary}.

\section{Minimal Kitaev chain in a dot-ABS pair} \label{sec:minimal_chain}
We first consider a minimal setup comprised of one quantum dot in the normal part and one ABS in the hybrid section. In particular, we derive the effective normal and Andreev couplings between them and the dependence of their ratio on experimentally accessible parameters.
Importantly, such a simple setup is sufficient for realizing a two-site Kitaev chain and can host poor man's Majorana zero modes at a fine-tuned sweet spot.

\subsection{Model Hamiltonian} \label{sec:model_hamiltonian}

The model Hamiltonian for a quantum dot-ABS pair is 
\begin{align}
& H_{DA} = H_D + H_A + H_T, \nn
& H_D = (\varepsilon_D + E_{ZD}) n_{D\su} + (\varepsilon_D - E_{ZD}) n_{D\sd} + U_D n_{D\su} n_{D\sd},  \nn
& H_A =(E_A + E_{ZA} ) \gamma^{\dagger}_{A\su} \gamma_{A\su}  + (E_A - E_{ZA}) \gamma^{\dagger}_{A\sd} \gamma_{A\sd}, \nn
&H_T = t_0 \sum_{\sigma, \eta=\uparrow,\downarrow}
c^{\dagger}_{\sigma} ( U_{so} )_{\sigma \eta} d_{\eta} + h.c..
\label{eq:H_DA}
\end{align}
Here $H_D$ is the Hamiltonian for a quantum dot with a single spinful orbital, which is a valid approximation when the dot level spacing is large.
$n_{D\sigma} = d\dg_{\sigma}d_{\sigma}$ is the occupancy number of the dot orbital with spin $\sigma$, $\varepsilon_D$ is the orbital energy, $E_{ZD}$ is the strength of the induced Zeeman spin splitting, and $U_D$ is the Coulomb energy.
$H_A$ is the Hamiltonian of the semiconductor-superconductor hybrid.
We assume that the low-energy physics of the hybrid is well described by a pair of subgap ABS, with all the above-gap continuum states being neglected.
$\gamma_{A\sigma} = \sigma u c_{\sigma} + v c\dg_{\overline{\sigma}}$ is the Bogoliubov operator of the ABS with $\sigma=\pm 1$ for spin $\su \sd$, and $u^2=1-v^2=1/2 + \varepsilon_A/2E_A$ are the BCS coherence factors.
$E_A=\sqrt{\varepsilon^2_A + \Delta^2_0}$ is the excitation energy, $\varepsilon_A$ is the normal-state energy, $\Delta_0$ is the induced pairing gap, and $E_{ZA}$ is the strength of the induced Zeeman spin splitting.
Here the Zeeman energy for both quantum dot and ABS are induced by the same globally applied magnetic field, and thereby the spin polarization axis of them coincide. 
However, owing to the $g$ factor renormalization at the semiconductor-superconductor interface~\cite{Stanescu2010Proximity, Reeg2018Metallization, Antipov2018Effects}, $E_{ZA}$ can be much weaker than $E_{ZD}$.
In our numerical simulations, we set $E_{ZA}= E_{ZD}/2$ without loss of generality.
$H_T$ is the tunnel Hamiltonian between dot and ABS, with $t_0$ being the tunneling amplitude which can be controlled by varying the tunnel gate voltage.
$U_{so}$ is a unitary matrix
\begin{align}
U_{so} &= e^{-i\alpha \sigma_{\theta}} \nn
&=\bpm
\cos \alpha - i \sin \alpha \sin \theta & -\sin \alpha \cos \theta\\
 \sin \alpha \cos \theta & \cos \alpha  + i \sin \alpha \sin \theta
\epm,
\label{eq:Uso}
\end{align}
with the diagonal and off-diagonal elements denoting the spin-conserving and spin-flipping processes, respectively.
Here $\alpha$ is the amount of spin precession accumulated in the tunnel region due to spin-orbit interaction.
$\sigma_{\theta} = \cos \theta \sigma_y +  \sin \theta \sigma_z$ is the spin-orbit field, which is perpendicular to the quantum dot chain axis.
Without loss of generality, here we have chosen a frame where the magnetic field direction and thus the dot spin polarization axis are fixed, and a rotation of the magnetic field is now equivalently described by rotating the spin-orbit field.
In particular, $\theta$ is the angle between the magnetic field and the dot chain axis, with $\theta=0$ ($\theta=\pi/2$) corresponding to being perpendicular (parallel) to the applied magnetic field.

\subsection{Low-energy effective theory and sweet spot conditions} \label{sec:low_energy}

We now derive the low-energy effective theory of the dot-ABS pair introduced in Eq.~\eqref{eq:H_DA}.
In the strong Zeeman field regime, the spin-down ABS state gets closer to the Fermi energy while the spin-up ABS becomes higher in energy and can be projected away in the leading-order approximation.
For the quantum dot, either of the spin-polarized orbitals can be closer to the Fermi energy than the other, depending on the value of the dot chemical potential.
Here, without loss of generality we restrict our discussions to the spin-down states as the low-energy degrees of freedom in both segments, leaving the discussions of other spin configurations in the supplemental material.
Therefore, in the weak tunneling, i.e., $t_0 \ll E_{ZD}, E_{ZA}$, the effective Hamiltonian of a dot-ABS pair is
\begin{align}
H^{\text{eff}}_{DA} &\equiv PH_{\text{DA}}P = (\varepsilon_D - E_{ZD}) d\dg_{\sd} d_{\sd} + (E_A - E_{ZA}) \gamma^{\dagger}_{A\sd} \gamma_{A\sd} \nn
& -t u \gamma\dg_{A\sd} d_{\sd} - t_{so} v \gamma_{A\sd} d_{\sd} + h.c.,
\label{eq:H_DA_eff}
\end{align}
where $P$ projects the original Hamiltonian onto the spin-down states, $t$ and $t_{so}$ are the tunnel amplitudes for the spin-conserving and spin-flipping processes, respectively, which are defined as
\begin{align}
    & t = (U_{so})_{\sd \sd} =t_0( \cos \alpha + i\sin \alpha \sin \theta ), \nn
    & t_{so} = (U_{so})_{\sd \su} =t_0\sin \alpha \cos \theta,
\label{eq:t_tso}
\end{align}
according to Eq.~\eqref{eq:Uso}.
Crucially, because an ABS is a coherent superposition of both electron ($u$) and hole ($v$) components, single electron tunneling from the quantum dot to the hybrid will simultaneously create and annihilate an ABS Bogoliubov excitation, giving both normal and Andreev-like effective couplings 
\begin{align}
\teff = -t u, \quad \Deff = -t_{so} v
\label{eq:t_D_eff}
\end{align}
between dot and ABS, as shown in Eq.~\eqref{eq:H_DA_eff}.
On the other hand, the Hamiltonian for a two-site Kitaev chain is
\begin{align}
H_{K2} = \varepsilon_1 f\dg_1 f_1 + \varepsilon_2 f\dg_2 f_2 + t_{12} f\dg_2f_1 + \Delta_{12} f_2 f_1 + h.c.,
\label{eq:H_K2}
\end{align}
where $f_i$ is the annihilation operator of a spinless fermion on site-$i$, $\varepsilon_i$ is the on-site energy, $t_{12}$ and $\Delta_{12}$ are the normal and Andreev-like tunneling between adjacent sites.
By comparing Eq.~\eqref{eq:H_DA_eff} with Eq.~\eqref{eq:H_K2}, we obtain the first main finding in the current work that the low-energy physics of a dot-ABS pair in the strong Zeeman regime is a two-site Kitaev chain.
In particular, the correspondence between the two is as below
\begin{align} \label{eq:kitaev_replacements}
    & f_1 \to d_{\sd}, \nn
    & f_2 \to \gamma_{A\sd}, \nn
    & \varepsilon_1 \to \varepsilon_D - E_{ZD}, \nn
    & \varepsilon_2 \to E_A - E_{ZA}, \nn
    & t_{12} \to \teff = -t u, \nn
    & \Delta_{12} \to \Deff =-t_{so} v.
\end{align}
Furthermore, the sweet spot of a two-site Kitaev chain is defined as $\varepsilon_1 = \varepsilon_2=0$ and $|t_{12}| = |\Delta_{12}|$.
That is, both the dot orbital energy 
\begin{align}
\varepsilon_D - E_{ZD} = 0,
\label{eq:condition1}
\end{align}
and the ABS energy
\begin{align}
\sqrt{\varepsilon^2_A + \Delta^2_0} - E_{ZA} = 0,
\label{eq:condition2}
\end{align}
need to be adjusted to be on resonance.
In addition, the magnitudes of normal and Andreev-like couplings need to be in perfect balance
\begin{align}
|tu|= |t_{so} v|.
\label{eq:condition3}
\end{align}
Once the sweet spot conditions indicated by Eqs.~\eqref{eq:condition1}-~\eqref{eq:condition3} are all satisfied, a pair of poor man's Majorana zero modes will emerge and localize themselves on the dot and hybrid segments, respectively, see also Fig.~\ref{fig:figure2} (d).

\subsection{Tuning Zeeman field strength}\label{sec:Zeeman_strength}

\begin{figure*}[t]
    \centering
    \includegraphics[width=\textwidth]{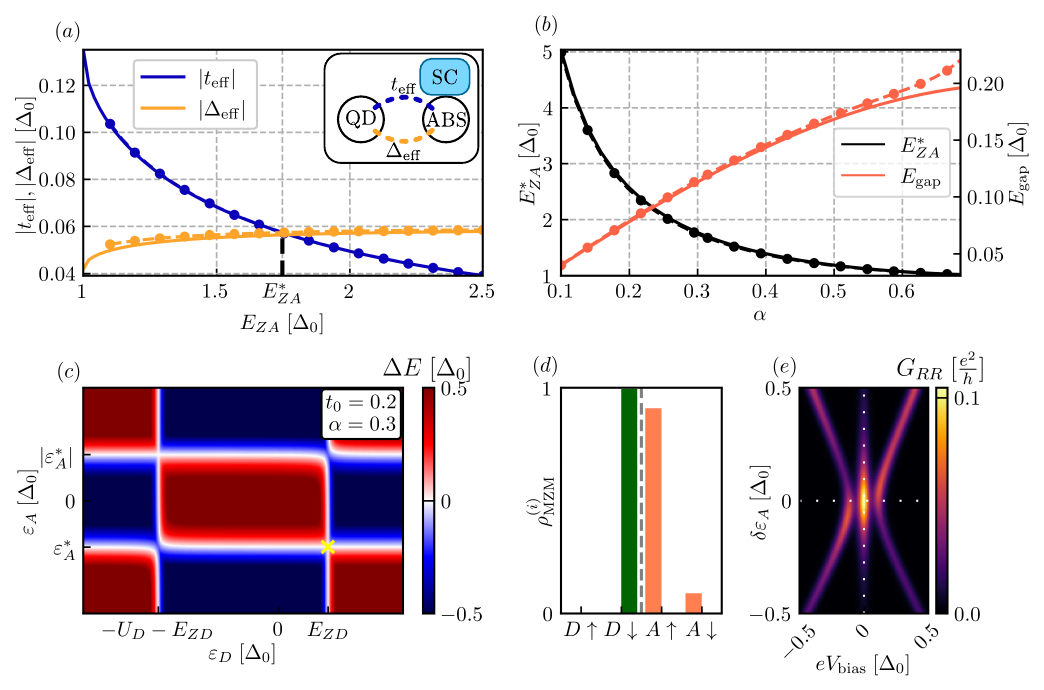}
    \caption{(a) strength of the effective couplings $\teff$ and $\Deff$ as a function of $E_{ZA}$ in a dot-ABS configuration. Analytic (numerical) results are presented with solid lines (dots). In the numerical calculations, the normal dot Zeeman energy is chosen to be $E_{ZD}=2E_{ZA}$. 
    At $E_{ZA} = E_{ZA}^*$, we find $|t_{\textrm{eff}}|=|\Delta_{\textrm{eff}}|$ . 
    (b) Sweet spot Zeeman energy $E_{ZA}^*$ and excitation gap $E_{\textrm{gap}}$ as a function of $\alpha$ due to spin-orbit interaction. 
    (c) Charge stability diagram at $E_{ZA}=E_{ZA}^*$. Here $\Delta E = E_{\text{odd, gs}} - E_{\text{even, gs}}$ is the energy difference between the ground states in opposite fermion parity subspaces. 
    The sweet spot is indicated by a yellow cross.
    (d) Majorana wavefunctions $\rho_{\mathrm{MZM}}^{(i)}$ at the sweet spot in (c). 
    (e) Local conductance $G_{RR}$ as a function of bias voltage $V_{\textrm{bias}}$ and ABS detuning $\delta \varepsilon_A$. }
    \label{fig:figure2}
\end{figure*}

We now consider how to reach the sweet spot in a dot-ABS pair by varying experimentally accessible parameters.
The most crucial step is the capability of tuning the relative amplitude of $\teff$ and $\Deff$.
In this subsection, we focus on using the Zeeman field strength as the tuning knob, which means one only varies the strength of the applied magnetic field, with its direction being fixed to be perpendicular to the Rashba spin-orbit field.
Setting $\theta=\pi/2$, we thereby have $t = t_0 \cos \alpha$ and $t_{so} = t_0 \sin \alpha$ in Eq.~\eqref{eq:t_tso}.
Among the three sweet spot conditions, the zero-energy dot orbital defined in Eq.~\eqref{eq:condition1} can be readily satisfied by merely varying the dot chemical potential.
By contrast, the other two defined in Eqs.~\eqref{eq:condition2} and~\eqref{eq:condition3} are more subtle and mutually constrained. 
Specifically, under a sufficiently large Zeeman field ($E_{ZA} > \Delta_0$), a zero-energy ABS is obtained only when the normal-state energy is pinned at
\begin{align}
\varepsilon_A^* \equiv - \sqrt{E^2_{ZA} - \Delta^2_0}<0.
\label{eq:var_A}
\end{align}
Note that here we particularly choose the negative $\varepsilon_A$ solution, corresponding to a hole-dominant ABS ($u < v$) such that a balance between $\teff$ and $\Deff$ indicated in Eq.~\eqref{eq:condition3} can be obtained in the weak spin-orbit interaction regime $t_{so} < t$ (a complete overview of all possible sweet-spot conditions is given in the Appendix). 
As a result, the dependence of the magnitudes of the effective couplings on the Zeeman field strength is as below
\begin{align}
& |\teff(E_{ZA})| = \frac{t_0 \cos \alpha}{\sqrt{2}}  \sqrt{ 1 - \sqrt{1 - \Delta^2_0/E^2_{ZA}}}, \nn
& |\Deff(E_{ZA})| = \frac{t_0  \sin \alpha }{\sqrt{2}} \sqrt{ 1 + \sqrt{1 - \Delta^2_0/E^2_{ZA}}}.
\label{eq:teff_Deff_EZA}
\end{align}
Equations~\eqref{eq:var_A} and~\eqref{eq:teff_Deff_EZA} show that increasing the Zeeman field strength, is changing the electron and hole components of the zero-energy ABS, i.e., increasing $v$ from $1/\sqrt{2}$ to close to 1, while decreasing $u$ from $1/\sqrt{2}$ to nearly zero.
Therefore, with $E_{ZA}$ increasing from $\Delta_0$, the Andreev coupling $|\Deff |$ is enhanced from $t_0 \sin \alpha /\sqrt{2}$ to $\sim t_0 \sin \alpha$ while the normal coupling $|\teff |$ is suppressed from $t_0 \cos \alpha/\sqrt{2}$ to zero in the large Zeeman limit [see Fig.~\ref{fig:figure2}(a)].
As a result, in the weak spin-orbit interaction regime ($\alpha < \pi/4$), which is experimentally relevant for InAs and InSb hybrid nanowires~\cite{Wang2022Singlet, Dvir2023Realization,Wang2023Triplet,Bordin2023Tunable}, the two coupling strengths will become equal at 
\begin{align}
E^*_{ZA} = \frac{\Delta_0}{\sin(2\alpha)},
\label{eq:E_ZA_star}
\end{align}
as indicated by the black dashed line in Fig.~\ref{fig:figure2}(a).
Furthermore, the excitation gap at this fine-tuned point is 
\begin{align}
E_{\text{gap}} = 2 |\teff(E^*_{ZA})| = t_0 \sin(2\alpha),
\label{eq:E_gap}
\end{align}
which is defined as twice the effective coupling strength.
As shown in Fig.~\ref{fig:figure2}(b), $E^*_{ZA}$ is a decreasing as a function of the spin-orbit interaction strength $\alpha$, while $E_{\text{gap}}$ is increasing.
In general, a larger $E^*_{ZA}$ is preferable in order to allow for a wider detuning range of the ABS energy $\delta \varepsilon_A \sim \sqrt{E^{*2}_{ZA} - \Delta^2_0}$.
Therefore, in choosing the optimal value of spin-orbit interaction $\alpha$ for the dot-ABS pair, there exists a tradeoff between a sizeable gap $E_{\text{gap}}$ and a large range of allowed $\delta \varepsilon_A$ for the effective Kitaev model.

To corroborate the analytic results obtained from the low-energy theory, we perform numerical simulations of the dot-ABS pair based on the full many-body Hamiltonian introduced in Eq.~\eqref{eq:H_DA}.
In particular, we choose $\Delta_0=1$ to be the natural unit, $U_D=5$, $t_0=0.2$ and $\alpha=0.3$, putting the system into the weak tunneling and weak spin-orbit interaction regime.
As shown in Fig.~\ref{fig:figure2}(a), the numerically calculated $|\teff |$ and $|\Deff |$ as a function of $E_{ZA}$ are in excellent agreement with the analytic results shown in Eq.~\eqref{eq:teff_Deff_EZA}.
In Fig.~\ref{fig:figure2}(b), the calculated $E^*_{ZA}$ and $E_{\text{gap}}$ also match very well with the analytical predictions in Eqs.~\eqref{eq:E_ZA_star} and~\eqref{eq:E_gap}.
Figure~\ref{fig:figure2}(c) shows the charge stability diagram in the $(\varepsilon_D, \varepsilon_A)$ plane.
A sweet spot, which is defined as the degeneracy point between even- and odd-parity ground states along with balanced normal and Andreev coupling strengths, appears in the right-bottom corner when $E_{ZA} \approx 1.75~\Delta_0$, consistent with the analytically predicted value of $E^*_{ZA}=1.77~\Delta_0$.
Here, the right-bottom corner corresponds to a spin-down dot orbital and a hole-dominant ABS, which is the focus of this section.
Furthermore, the calculated wavefunctions in Fig.~\ref{fig:figure2}(d) show that indeed, a pair of Majorana zero modes emerge at the sweet spot [yellow cross in Fig.~\ref{fig:figure2}(c)], localized at the quantum dot and hybrid, respectively.
In Fig.~\ref{fig:figure2}(e), the calculated tunnel conductance spectroscopy in the $(V_{\text{bias}}, \delta \varepsilon_A)$ plane shows a stable zero-bias peak and a parabola-shaped gap peak, consistent with the conductance features of poor man's Majorana zero modes.

\subsection{Tuning Zeeman field direction} \label{sec:direction_of_zeeman}
\begin{figure}
    \centering
    \includegraphics[width=\linewidth]{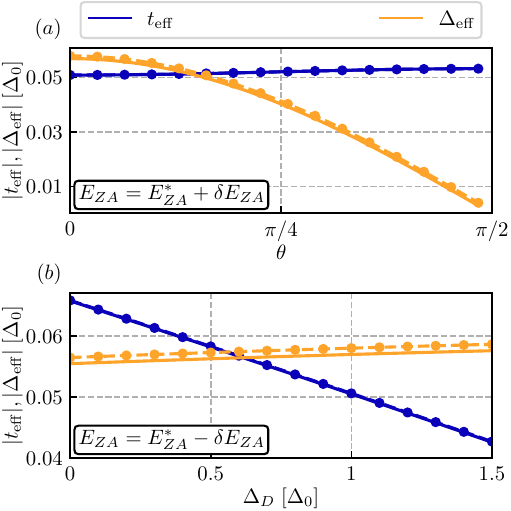}
    \caption{
    (a) Strength of the effective couplings $|\teff|$ and $|\Deff|$ as a function of the field angle $\theta$ at $E_{ZA} = E^*_{ZA} + \delta E_{ZA}$ ($\delta E_{ZA}=0.2$).
    Here $\theta$ is the angle between the magnetic field and the dot chain axis, and in particular, $\theta=0$ corresponds to the magnetic field being perpendicular to the spin-orbit field.
    (b) $|\teff|$ and $|\Deff|$ as a function of induced gap $\Delta_D$ at $E_{ZA} = E^*_{ZA} - \delta E_{ZA}$.
    In both scenarios, a perfect balance between $|\teff|$ and $|\Deff|$ can be obtained.}
    \label{fig:figure3}
\end{figure}

We now consider rotating the applied magnetic field in order to find the sweet spot, with the field strength being fixed.
Inside the rotation plane, the field direction can be either parallel or perpendicular to the spin-orbit field~\cite{Bommer2019SpinOrbit, Liu2019Conductance}.
In our consideration, this field rotation is equivalently described by rotating the spin-orbit field while fixing the Zeeman field and spin polarization axis, as explained after Eq.~\eqref{eq:Uso}.
While increasing field strength changes $u$ and $v$ of the zero-energy ABS, the effect of field rotation is to change the ratio of the spin-conserving $t$ and spin-flipping amplitudes $t_{so}$, as indicated in Eq.~\eqref{eq:t_tso}.
Plugging Eq.~\eqref{eq:t_tso} into Eq.~\eqref{eq:t_D_eff}, we thus obtain
\begin{align}
& |\teff(\theta)| = t_0 \sqrt{1 - \sin^2\alpha \cos^2 \theta} \cdot u(E_{ZA}), \nn
& |\Deff(\theta)| = t_0 \sin \alpha \cos \theta \cdot v(E_{ZA}),
\end{align}
where $u$ and $v$ do not depend on angle $\theta$.
Here we only focus on $0 \leq \theta \leq \pi/2$, since the strength of the effective couplings are $\pi$-periodic and symmetric about $\theta=0$. 
As shown in Fig.~\ref{fig:figure3}(a), $|\teff|$ ($|\Deff|$) is an increasing (decreasing) function of the field angle $\theta$.
In particular, when the magnetic field aligns with the spin-orbit field ($\theta=\pi/2$), $|\Deff|$, which is of triplet nature, is suppressed completely due to spin conservation.
In order to obtain a sweet spot in the angle sweep, one thereby needs to start with a sufficiently strong Zeeman field ($E_{ZA} > E^*_{ZA}$), giving $|\Deff| > |\teff| $ at $\theta=0$, and then rotate the magnetic field to reach the balance between $|\Deff|$ and $|\teff|$.
Thus, in general a larger excitation gap would appear in the vicinity of $\theta=0$, where the spin-flipping processes are maximized.

\subsection{Tuning induced pairing gap in the quantum dot} \label{sec:proximity_dot}
The third tuning knob we consider in the current work is the superconducting pairing gap in the normal quantum dot, which can be induced from the adjacent hybrid by proximity effect.
Microscopically, this proximity effect can originate from either the ABS or the continuum states, with the forms being
\begin{align}
    & \Delta_{\text{ABS}} = (t^2 + t^2_{so}) \frac{uv}{E_A + E_{ZA}}, \nn
    & \Delta_{\text{cont}} = (t^2 + t^2_{so}) \frac{\chi}{\Delta_{0}},
\end{align}
up to the leading order.
Here $\Delta_{\text{ABS}}$ comes from the high-energy spin-up ABS, while $\Delta_{\text{cont}}$ is obtained assuming a zero-band-width model for the continuum states, with $\chi$ characterizing the continuum density of states which can be quite different from the ABS.
Since both $\Delta_{\text{ABS}}$ and $\Delta_{\text{cont}}$ increase with the tunnel amplitude $t_0$, their strength can be experimentally enhanced by lowering the tunnel barrier between dot and hybrid.
In the following calculations, we do not distinguish between the microscopic origins of the proximity effect, instead we consider their combined effect in a phenomenological way (examples of a microscopic model to change the induced superconductivity by tuning $t_0$ are given in the Appendix).
Now the dot Hamiltonian becomes
\begin{align}
& H'_D = H_D + H_{\text{ind}}, \nn
& H_{\text{ind}} = \Delta_D d\dg_{\su} d\dg_{\sd} +h.c.,
\end{align}
where $H_D$ is the bare dot Hamiltonian defined in Eq.~\eqref{eq:H_DA}, and $\Delta_D$ is the total induced gap on the dot.
As a result, the dot orbital in the quantum dot is now proximitized into an Yu-Shiba-Rusinov state~\cite{Yu1965Bound, Shiba1968Classical, Rusinov1969Theory, Rasmussen2018YSR}, with the electron and hole components being 
\begin{align}
 u_D &= \sqrt{\half + \frac{\xi_D}{2(U_D/2 + E_{ZD})}} \approx 1 - \frac{1}{8} \left( \frac{\Delta_D}{  U_D/2 + E_{ZD} } \right)^2,\nn
 v_D &= \sqrt{\half - \frac{\xi_D}{2(U_D/2 + E_{ZD})}} \approx \half \left( \frac{\Delta_D}{ U_D/2 + E_{ZD} } \right),
 \label{eq:uD_vD}
\end{align}
where 
\begin{align}
\xi_D \equiv \varepsilon_D + U_D/2 = \sqrt{\left( \frac{U_D}{2} + E_{ZD}\right)^2 - \Delta^2_D}
\end{align}
is determined by the zero-energy condition for the bound state.
Note that the approximations in Eq.~\eqref{eq:uD_vD} are made in the weak proximity effect regime ($\Delta_D \ll U_D$), and thereby, up to the leading order of $\Delta_D/U_D$, $u_D=1$ becomes a constant and only $v_D \propto \Delta_D$ grows linearly with $\Delta_D$.
As a result, the effective couplings between dot and ABS become
\begin{align}
& \teff = t( u_A u_D - v_A v_D ) \approx t( u_A - v_A v_D ) , \nn
& \Deff = t_{so} (v_A u_D + u_A v_D ) \approx t_{so} (v_A + u_A v_D ).
\label{eq:teff_Deff_DeltaD}
\end{align}
That is, $\teff$ decreases with the magnitude of the induced pairing, while $\Deff$ increases with it.
In Fig.~\ref{fig:figure3}(b), the solid lines show the analytic curves of $|\teff|$ and $|\Deff|$ as a function of $\Delta_D$ derived in Eq.~\eqref{eq:teff_Deff_DeltaD}, which agree with the numerical results obtained from the full many-body Hamiltonian (dots and dashed lines).
Note that here the Zeeman field is perpendicular to the spin-orbit field, and its strength is chosen to be $E_{ZA} < E^*_{ZA}$ such that $|\teff| > |\Deff|$ at zero proximity effect, and a balance between them is reached at a sufficiently strong $\Delta_D$.

\section{Scaling up the Kitaev chain} \label{sec:scaling_up}
We now go beyond the minimal setup of a dot-ABS pair and scale up the system into a longer chain.
Without loss of generality, we consider tuning up the sweet spot and Majorana modes by varying the Zeeman field strength, with its direction being fixed to be perpendicular to the spin-orbit field and no superconducting proximity effect on normal quantum dots.
Moreover, we assume homogeneity in the long-chain system, i.e., all the physical parameters for the dots/ABS/tunneling are identical.

\subsection{Three-site Kitaev chain: dot-ABS-dot}

\begin{figure}
    \centering
    \includegraphics[width=\linewidth]{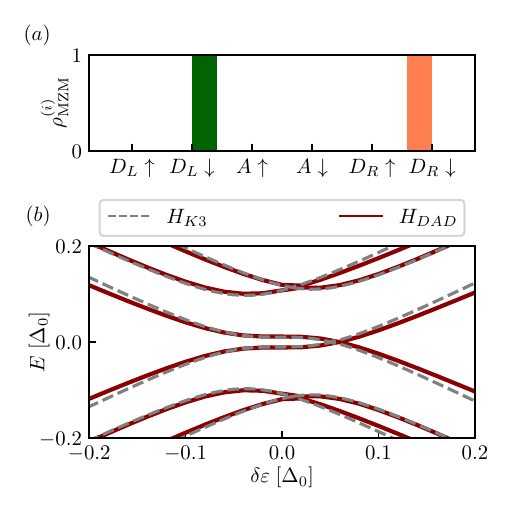}
    \caption{
    (a) Majorana wavefunctions at the sweet spot of a dot-ABS-dot chain.
    (b) Energy spectrum of the system as a function of detuning energy $\delta\varepsilon$. 
    The solid lines are calculated for the full many-body Hamiltonian, while the grey dashed lines are based on the low-energy effective theory.
    The energy splitting at $\delta \varepsilon=0$ is due to effective next-nearest-neighbor coupling between the two outer dots.
    }
    \label{fig:figure4}
\end{figure}

As a first example of the three-site Kitaev chain, we consider a dot-ABS-dot chain, focusing on its physical properties around the sweet spot.
The Hamiltonian is given by
\begin{align}
H_{\text{DAD}} = H_{DL} + H_{A} + H_{DR} + H_{TLA} + H_{TRA},
\label{eq:H_DAD}
\end{align}
where $H_{DL}, H_{A}$ and $H_{DR}$ are the Hamiltonians for the left dot, the middle ABS, and the right dot, respectively.
$H_{TLA}$ ($H_{TRA}$) is tunnel Hamiltonian between the ABS and the left (right) dot.
The specific forms of these individual Hamiltonian terms are the same as those introduced in Eq.~\eqref{eq:H_DA}.
Under the assumption of homogeneity, one can simultaneously tune both dot-ABS pairs into their own sweet spot by applying a global Zeeman field $E_{ZA}= E^*_{ZA}$ and setting $\varepsilon_{DL} = \varepsilon_{DR} = \varepsilon^*_D$ and  $\varepsilon_A = \varepsilon^*_A$ as indicated in Fig.~\ref{fig:figure2}(c), such that the whole system is automatically entering the sweet spot regime.
Indeed, as shown in Fig.~\ref{fig:figure4}(a), two unpaired Majorana modes are completely localized on the outermost quantum dots, precisely as expected for the sweet spot of a three-site Kitaev chain~\cite{Kitaev2001Unpaired}.
However, a surprising fact is that the energy of the two Majoranas are split into $E_{MZM} \approx 0.01$ which is approximately one-tenth of the excitation gap, [see Fig.~\ref{fig:figure4}(b) at $\delta \varepsilon=0$], even though there is no wavefunction overlap between them at all [see Fig.~\ref{fig:figure4}(a)].
Furthermore, as shown in Fig.~\ref{fig:figure4}(b), the energy spectrum of the whole system as a function of the detuning energy deviates from the cubic scaling behavior $E \propto (\delta \varepsilon)^3$ of an idealized three-site Kitaev chain.
Here the detuning energy is defined as $\delta \varepsilon = \varepsilon_{DL} - \varepsilon^*_{D} = \varepsilon_{DR} - \varepsilon^*_{D} = \varepsilon_{A} - \varepsilon^*_A$.

\begin{figure}[t]
    \centering
    \includegraphics{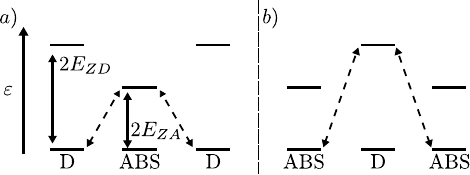}
    \caption{
    Schematic of the second-order tunneling processes that are responsible for the next-nearest-neighbor couplings in a dot-ABS-dot or ABS-dot-ABS chain.
    }
    \label{fig:figure5}
\end{figure}

To understand the physical mechanism underlying this intriguing energy splitting, we develop a low-energy effective theory for the dot-ABS-dot chain, including both the first- and second-order contributions,
\begin{align}
H_{\text{DAD,eff}} = H^{(1)}_{\text{DAD,eff}}  + H^{(2)}_{\text{DAD,eff}} .
\label{eq:H_DAD_eff}
\end{align}
Here $H^{(1)}_{\text{DAD,eff}}$ includes only the low-energy states and direct tunneling terms, which is a straightforward generalization of the dot-ABS pair, giving
\begin{align}
& H^{(1)}_{\text{DAD,eff}}=PH_{\text{DAD}}P \nn
&= (\varepsilon_{DL} - E_{ZD}) d\dg_{L\sd} d_{L\sd} + (E_{A} - E_{ZA}) \gamma^{\dagger}_{A\sd} \gamma_{A\sd} \nn
& + (\varepsilon_{DR} - E_{ZD}) d\dg_{R\sd} d_{R\sd} 
+ t_{\text{eff}LA} \gamma\dg_{A\sd} d_{L\sd} + \Delta_{\text{eff}LA} \gamma_{A\sd} d_{L\sd} \nn
& + t_{\text{eff}RA} \gamma\dg_{A\sd} d_{R\sd} + \Delta_{\text{eff}RA}  \gamma_{A\sd} d_{R\sd}
+ h.c..
\label{eq:H_DAD_1}
\end{align}
Indeed, the first-order effective Hamiltonian $H^{(1)}_{\text{DAD,eff}}$ is a three-site Kitaev chain.
In particular, the sweet spot is reached when all the onsite energies are zero and $E_{ZA} = E^*_{ZA}$ giving $t_{\text{eff}LA} = t_{\text{eff}RA} = \Delta_{\text{eff}LA} = \Delta_{\text{eff}RA}$.
In addition, unlike the dot-ABS pair, we now also include the second-order perturbation terms into the effective Hamiltonian as below
\begin{align}
H^{(2)}_{\text{DAD,eff}} &= P H_T \frac{1-P}{H_A} H_T P \nn
&= t_{DD} d\dg_{L\sd} d_{R\sd} + 
\Delta_{DD} d_{L\sd} d_{R\sd} + h.c.,
\label{eq:H_2_eff}
\end{align}
where $H_T = H_{TLA} + H_{TRA}$.
Equation~\eqref{eq:H_2_eff} indicates that effective next-nearest-neighbor couplings between the outer dots can be mediated by the high-energy ABS in the hybrid via second-order tunnelings (see Fig.~\ref{fig:figure5}).
Specifically, these couplings have the following form
\begin{align}
& t_{DD} = \frac{t^2 v^2 + t^2_{so} u^2 }{ 2E_{ZA} } \approx \frac{t^2 v^2 }{ 2E_{ZA} }, \nn
& \Delta_{DD} = \frac{2tt_{so} uv }{2E_{ZA}} \ll t_{DD},
\label{eq:t_DD}
\end{align}
where we assume the weak spin-orbit limit $t_{so}\ll t$ and $u \ll v$ holds in the vicinity of the sweet spot.
Therefore, up to the second order in $t_0$, the low-energy physics of a dot-ABS-dot chain is well described by a generalized three-site Kitaev chain
\begin{align} 
H_{K3} = &\sum^3_{i=1} \varepsilon_i f\dg_i f_i + \sum^2_{i=1} ( t f\dg_{i+1} f_i + \Delta f_{i+1} f_i ) \nn
&+ t_{31} f\dg_3 f_1 + \Delta_{31} f_3 f_1 + h.c.,
\label{eq:h_k3}
\end{align}
where $f_i$ is the spinless fermion on the $i$-th site, $\varepsilon_i$ is the on-site energy, $t$ and $\Delta$ are the normal and Andreev tunnelings between adjacent sites, and $t_{31}$ and $\Delta_{31}$ are the next-nearest-neighbor tunnelings.
Indeed, as shown in Fig.~\ref{fig:figure4}(b), the energy spectrum of the full many-body Hamiltonian in Eq.~\eqref{eq:H_DAD} is in excellent agreement with that of the effective model in Eq.~\eqref{eq:h_k3}, supporting our perturbation theory analysis.
In the calculation of the generalized Kitaev model, the Hamiltonian parameters are chosen as $t=\Delta =t_{\text{eff}}(E^*_{ZA}), t_{31} = t_{DD}, \Delta_{31}=0$ and $\varepsilon_i = \delta \varepsilon$.
Therefore our new finding here is that even though the two Majorana modes have no wavefunction overlap in space at the sweet spot [see Fig.~\ref{fig:figure4}(a)], they are still coupled to each other via next-nearest-neighbor couplings, giving a finite energy splitting [see Fig.~\ref{fig:figure4}(b)].
In App. \ref{app:inhom} we expand our discussion to the case of inhomogeneities of g-factors and spin-orbit mixing between the constituing dots of the array.

\subsection{Three-site Kitaev chain: ABS-dot-ABS}

\begin{figure}
    \centering
    \includegraphics[width=\linewidth]{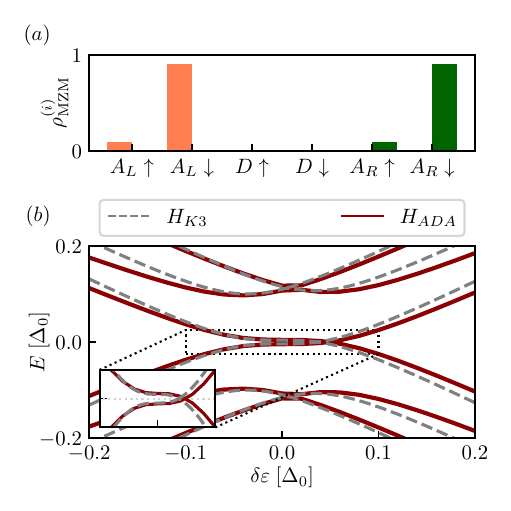}
    \caption{
    (a) Majorana wavefunctions at the sweet spot of an ABS-dot-ABS chain.
    (b) Energy spectrum of the system as a function of detuning energy $\delta\varepsilon$. 
    The solid lines are calculated for the full many-body Hamiltonian, while the grey dashed lines are based on the low-energy effective theory.
    The energy splitting at $\delta \varepsilon=0$ is due to effective next-nearest-neighbor coupling between the two outer ABS.
    }
    \label{fig:figure6}
\end{figure}

We briefly discuss the physics of an ABS-dot-ABS chain which is somewhat dual to a dot-ABS-dot chain.
The Hamiltonian is given by
\begin{align}
H_{\text{ADA}} = H_{AL} + H_{D} + H_{AR} + H_{TLD} + H_{TRD},
\label{eq:H_ADA}
\end{align}
which has two outer ABS connected by a quantum dot in the middle.
Similar to the analysis performed in the previous subsection, the low-energy physics of the system is
\begin{align}
H_{\text{ADA,eff}} =H^{(1)}_{\text{ADA,eff}}  + H^{(2)}_{\text{ADA,eff}},
\end{align}
where the first-order term is 
\begin{align}
& H^{(1)}_{\text{ADA,eff}} = (E_{AL} - E_{ZA}) \gamma\dg_{L\sd} \gamma_{L\sd} + (\varepsilon_{D} - E_{ZD}) d^{\dagger}_{\sd} d_{\sd} \nn
& + (E_{AR} - E_{ZA}) \gamma\dg_{R\sd} \gamma_{R\sd} 
+ t_{\text{eff}LD} d\dg_{\sd} \gamma_{L\sd} + \Delta_{\text{eff}LD} d_{\sd} \gamma_{L\sd} \nn
&+ t_{\text{eff}RD} d\dg_{\sd} \gamma_{R\sd} + \Delta_{\text{eff}RD} d_{\sd} \gamma_{R\sd} + h.c.,
\label{eq:H_ADA_1}
\end{align}
and the second-order term is 
\begin{align}
H^{(2)}_{\text{ADA,eff}} = t_{AA} \gamma\dg_{L\sd} \gamma_{R\sd} + \Delta_{AA} \gamma_{L\sd} \gamma_{R\sd} + h.c.,
\label{eq:H_ADA_2}
\end{align}
with
\begin{align}
& t_{AA} = \frac{t^2 v^2 + t^2_{so} u^2 }{ 2E_{ZD} } \approx \frac{t^2 v^2}{ 2E_{ZD} }, \nn 
& \Delta_{AA} = \frac{2tt_{so} uv }{2E_{ZD}} \ll  t_{AA}.
\label{eq:t_AA}
\end{align}
We thus see that the low-energy physics of a ABS-dot-ABS chain is also a generalized three-site Kitaev chain, with only the roles of quantum dots and ABS being interchanged.
Actually the sweet spot of the system is also reached at $E_{ZA}=E^*_{ZA}$, giving an excitation energy gap of similar size with its dual system.
The only difference is a more suppressed Majorana energy splitting [see Fig.~\ref{fig:figure6}(b)] because a larger Zeeman spin splitting in the quantum dot suppresses the second-order tunnelings, as indicated in Eq.~\eqref{eq:t_AA}.
As for the dot - ABS - dot setup, in App. \ref{app:inhom} we expand the discussion to situations where the dots have inhomogeneities in either g-factors or spin-orbit mixing. 
In particular we find that the ABS - dot - ABS system yields more resillient sweet spots due to the larger separation of levels on the central normal dot.

\begin{figure*}[t]
    \centering
    \includegraphics[width=\linewidth]{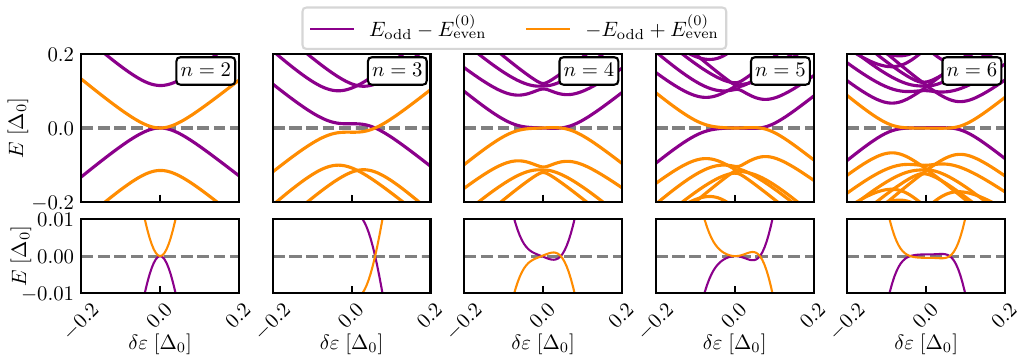}
    \caption{Energy spectra of $N$-site alternating quantum dot-ABS chains around the sweet spot.
    For intermediate length chains (e.g., $N=3$), the next-nearest-neighbor coupling has an appreciable effect on the Majorana energy splitting at the sweet spot. 
    As the chain is further scaled up ($N \geq 4$), this Majorana energy splitting is quickly suppressed due to the short-range nature of this effective coupling as shown in Eq.~\eqref{eq:gamma_k}.}
    \label{fig:figure7}
\end{figure*}

\subsection{Longer Kitaev chain}
For a general Kitaev chain with $N$ alternating quantum dots and ABS in total, its low-energy physics can be well described by an effective Hamiltonian up to the $N-1$-th order, i.e.,
\begin{align}
    H_{N,\text{eff}} = \sum^{N-1}_{k=1} H^{(k)}_{N,\text{eff}}.
\end{align}
In particular, the strength of the effective couplings between two arbitrary sites in such an $N$-site chain has the following scaling behavior
\begin{align}
\Gamma_{k} \sim \frac{t^k_0}{(2E_Z)^{k-1}} \sim t_0 \exp \{ -(k-1)\log ( 2E_Z/t_0 ) \},
\label{eq:gamma_k}
\end{align}
for $1 \leq k \leq N-1$.
Here $\Gamma_k$ denotes the effective normal or Andreev coupling between any two sites separated by a distance of $k$ (e.g., $\Gamma_1$ for the nearest-neighbor coupling), and $E_Z$ is the Zeeman spin splitting of either dot or ABS located between the two sites considered.
Physically, $\Gamma_k$ originates from virtual tunnelings that include $k$ times of single-electron tunneling events via $k-1$ different high-energy states that are gapped from the Fermi energy due to Zeeman spin splitting.
From Eq.~\eqref{eq:gamma_k}, we see that $\Gamma_k$ decays exponentially with the distance between the coupled two sites, with the decay length being approximately
\begin{align}
\xi^{-1}_{\Gamma} \sim \log(2E_Z/t_0).
\label{eq:xi}
\end{align}
The range of the effective couplings thus decreases with an increase of Zeeman energy in the dot-ABS chain, and as a result, the low-energy physics of the dot-ABS chain will asymptotically approach the idealized spinless Kitaev chain only in the large Zeeman energy limit ($E_Z \gg t_0$).
Another crucial feature of $\Gamma_{k}$ is being short-ranged in nature, making it possible to reach topological protection in the long-chain limit.
Indeed, the numerical simulations based on the full many-body Hamiltonian for $N$-site dot-ABS chain (see Fig.~\ref{fig:figure7}) show that the finite Majorana energy splitting become strongly suppressed to nearly zero once there are as many as four or five sites.
Moreover, for $N$ as large as six, the range of $\delta \varepsilon$ for hosting zero-energy excitation extends asymptotically to the long-wire limit $ -t_{\text{eff}}(E^*_{ZA}) \lesssim \delta \varepsilon \lesssim t_{\text{eff}}(E^*_{ZA})$ and signatures of gap closing and reopening begin to appear near $|\delta \varepsilon| \sim t_{\text{eff}}$, indicating the emergence of topological phase transition.

\section{Discussion} \label{sec:discussion}

In the current work, we proposed a new way of implementing a Kitaev chain in an alternating quantum dot-Andreev bound state array.
Although the configuration of the proposed hybrid devices resembles those considered in Refs.~\cite{Sau2012Realizing, Leijnse2012Parity, Liu2022Tunable, Tsintzis2022Creating}, a fundamental difference is the role of ABS.
In Refs.~\cite{Liu2022Tunable, Tsintzis2022Creating}, the ABSs are \emph{gapped} and only serve as a virtual coupler to mediate the effective couplings between quantum dots.
By contrast, here the spin-polarized ABS is taken close to the Fermi energy and are on equal footing with the dot orbitals as the spinless fermions.
Consequently, an immediate advantage of our proposal is to emulate a Kitaev chain using a reduced number of quantum dots and hybrid segments in a device.
In particular, it becomes possible to implement a two-site Kitaev chain and poor man's Majoranas using only one quantum dot and one hybrid segment.
Furthermore, the existing two-site Kitaev chain device comprised of a double quantum dot linked by a hybrid~\cite{Dvir2023Realization} is now suitable for realizing a three-site Kitaev chain exhibiting the physics of bulk-edge correspondence in the vicinity of its sweet spot.
On the other hand, our proposal differs from the ABS chain proposed in Ref.~\cite{Fulga2013Adaptive} in that we require only half the number of superconducting leads and that we do not need to control the quantum point contact between the semiconductor wire and the superconductor leads, making our theoretical proposal more experimentally accessible.
Another advantage of our proposal is the ability of getting a relatively large excitation gap, because now the effective couplings originate from direct couplings of the dot-ABS pair, i. e., $E_{\text{gap}} \sim t_0 $ in stark contrast with the second-order tunneling processes $E_{\text{gap}} \sim t^2_0/\Delta_0 $ in the previous works~\cite{Liu2022Tunable}

Throughout the work, we have assumed perfect homogeneity when considering a long Kitaev chain device ($N \geq 3$), but this has to be relaxed in a realistic device.
That is, the quantum dots can have different values of charging energy $U_D$, $g$ factor $E_{ZD}$, while $E_{ZA}$ and induced gap $\Delta_0$ of the ABS may vary from piece to piece.
Therefore, it would be rather unlikely to drive the whole long chain into the sweet spot by merely controlling a global magnetic field, and as a result, the tuning knob of induced gap on quantum dots become particularly crucial, because it will allow for fine-tuning the couplings in each individual dot-ABS pair into perfect balance.

Another new finding of our work is the presence of couplings beyond the nearest neighbors, which originate from high-order tunneling processes.
Its effect will be most prominent in a three-site device (e.g., dot-ABS-dot chain), where the Majorana energy at the sweet spot becomes split even though their wavefunctions are completely separated on the outermost dots.
This raises a new open question of whether it is possible to define such a sweet spot which simultaneously satisfies three conditions: 1. complete spatial separation of the Majoranas, 2. robustness against onsite-energy detuning, and 3. minimizing the Majorana energy to nearly zero.
In order to obtain an idealized Kitaev chain model, as shown in Eqs.~\eqref{eq:t_DD} and~\eqref{eq:t_AA}, the couplings between two distant sites would be suppressed in the strong Zeeman limit, similar to the findings in the Majorana nanowire scenarios~\cite{Pan2023Majorana}.
In addition, in the tunneling regime $t_0 \ll E_Z$, such couplings are short-ranged in nature, and therefore the effect will be mitigated as the number of sites is scaled up.
As we show, when the number of sites is as large as six, the whole chain becomes very close to a topological Kitaev chain with robust zero energy and signatures of gap closing and reopening near the quantum phase transition.

\section{Summary} \label{sec:summary}
To summarize, we have proposed a new route to simulating Kitaev chain in an alternating quantum dot-Andreev bound state array.
In particular, both the dot orbitals and the ABS are now on equal footing as spinless fermions, and the relative amplitude of normal and Andreev couplings between adjacent sites are highly tunable by the strength and direction of the magnetic field, as well as the magnitude of the induced pairing gap on quantum dots.
As the quantum dot-ABS chain is scaled up, couplings beyond the nearest neighbors emerge, affecting the Majorana energy at the sweet spot.
Nonetheless, due to short-range nature of these couplings, topological protection of Majorana zero modes will recover in the long chain limit.
Our proposal will allow for a more efficient simulation of artificial Kitaev chain using a reduced number of quantum dot or hybrid segment and will at the same time obtain a larger excitation gap above the Majorana zero modes.
In recent experiments~\cite{Bordin2023Tunable, Zatelli2023Robust}, it has been demonstrated possible to isolate a single ABS in a short hybrid region, making our proposal particularly appealing and relevant to the ongoing studies.
Finally, we have been made aware of \cite{Samuelson2023minimal} by its authors. 
In contrast to \cite{Samuelson2023minimal}, we consider the the limit of both quantum dots being maximally asymmetric in their proximity coupling to the superconductor.

\begin{acknowledgements}
We are particularly grateful to S. L. D. ten Haaf, and B. Roovers for sharing and discussing experimental data with us, motivating this work.
We further thank A. M. Bozkurt, J. D. Torres Luna, and K. Vilkelis for useful theory discussions.
We are grateful to the authors of \cite{Samuelson2023minimal} for making us aware of their work prior to publication.
This work was supported by a subsidy for top consortia for knowledge and innovation (TKI toeslag), by the Dutch Organization for Scientific Research (NWO). S.M. acknowledges funding of NWO through OCENW.GROOT.2019.004.
\end{acknowledgements}

\textit{Author contributions.}-- 
C.-X.L. formulated the project idea and designed the project with input from D.v.D and M.W.
S.M. and C.-X.L. performed the calculations.
S.M. generated the figures with input from C.-X.L.
C.-X.L. and M.W. supervised the project.
All authors discussed the results and contributed to writing of the manuscript.

\textit{Data availability.}--
All codes related to the results of the present manuscript can be found in the accompanying Zenodo repository \cite{DotABSZenodo}.

\bibliography{references.bib}

\appendix

\onecolumngrid
\vspace{1cm}
\begin{center}
{\bf\large Appendix}
\end{center}
\vspace{0.5cm}

\setcounter{secnumdepth}{3}
\setcounter{equation}{0}
\setcounter{figure}{0}
\renewcommand{\theequation}{S-\arabic{equation}}
\renewcommand{\thefigure}{S\arabic{figure}}
\renewcommand\figurename{Supplementary Figure}
\renewcommand\tablename{Supplementary Table}
\newcommand\Scite[1]{[S\citealp{#1}]}
\newcommand\Scit[1]{S\citealp{#1}}

\makeatletter \renewcommand\@biblabel[1]{[S#1]} \makeatother

\section{Details of the numerical calculations}\label{app:numerical_details}

The numerical results in this work are obtained by exact diagonalization of the full many-body Hamiltonian, e.g., Eq.~\eqref{eq:H_DA}, in Fock space.
The dimension of the total Hamiltonian is $2^{2N}$ where $N$ is the number of quantum dots plus ABS in the chain.
Due to fermion parity conservation, we can decompose the Hamiltonian into even- and odd-parity subspace of dimension $2^{2N-1}$.
As a result, the excitation energy $E$, as depicted in Figs.~\ref{fig:figure4}, \ref{fig:figure6}, and \ref{fig:figure7}, has been obtained through
\begin{align} \label{eq:numerical_energy}
    E_{\lambda}=E_{\mathrm{odd}}^{(\lambda)}-E_{\mathrm{even}}^{(0)}
\end{align}
for $\lambda = 0, 1, 2, ..., 2^{2N-1}-1$.
In a similar fashion, $\Delta E$ in the charge stability diagrams, as depicted in Fig.~\ref{fig:figure2}(c), is obtained by restricting $\lambda$ in Eq.~\eqref{eq:numerical_energy} to $\lambda=0$. 
To obtain $t_{\mathrm{eff}}$ and $\Delta_{\mathrm{eff}}$ we make use of the following relation
\begin{align}
& \Deff = (E_1 + E_0)/2, \nn
& \teff = (E_1 - E_0)/2, 
\end{align}
where $E_0, E_1$ are defined in Eq.~\eqref{eq:numerical_energy}.

Here $t_{\mathrm{eff}}$ is related to the cost of exciting an unpaired electron while $\Delta_{\mathrm{eff}}$ is related to splitting the lowest Cooper pair into two unpaired electrons. 
The gap energy, depicted in e.g. Fig.~\ref{fig:figure2}(b), is obtained directly through
\begin{align}
    E_{\mathrm{gap}} \equiv E_1 .
\end{align}
In addition, the spectrum plots also contain the energy of the MZM itself in the lowest laying line following 
\begin{align}
    E_{\mathrm{MZM}}\equiv E_0.
\end{align}
The Zeeman energies characterising the sweet spot depicted in Fig.~\ref{fig:figure2} b) have been determined by using the properties of the CSD [see Fig.\ref{fig:figure2} (c)].
As Fig.~\ref{fig:figure3} shows, if $E_{ZA}$ is slightly below the sweet spot $E_{ZA}^*$, the degeneracy crossing vanishes in favor of an $t_{\mathrm{eff}}$ dominated ($\Delta E>0$) anti-crossing.
For $E_{ZA}>E_{ZA}^*$ the same anti-crossing is caused by a dominating $\Delta_{\mathrm{eff}}$ process ($\Delta E<0$).
Therefore, the sweet spot is characterised by the root $\Delta E(E_{ZA}^*)=0$ at the point where the charge degeneracy lines have their smallest distance in the $(\varepsilon_D,\varepsilon_A)$ plane.
To determine the point of minimal distance, we perform a transformation of the chemical potentials into polar coordinates
\begin{align}
    (\varepsilon_D, \varepsilon_A) \rightarrow (r_\varepsilon \cos(\varphi_\varepsilon), r_\varepsilon \sin(\varphi_\varepsilon)) \quad r_\varepsilon=\sqrt{\varepsilon_D^2+\varepsilon_A^2}, \; \tan(\varphi_\varepsilon)=\varepsilon_A/\varepsilon_D.
\end{align}
In polar coordinates, the point with minimal distance between the degeneracy lines is found to satisfy 
\begin{align}
    (r_\varepsilon, \varphi_\varepsilon)\in \{r_\varepsilon,\varphi_\varepsilon :\mathrm{min}[\Delta E(r_\varepsilon,\varphi_\varepsilon)] \wedge \mathrm{max}[\Delta E(r_\varepsilon, \varphi_\varepsilon)]\}.
\end{align}
The bottom right corner we discuss has the further constraint that $\varphi_\varepsilon\in[-\pi/2, 0]$.
To obtain the MZM wavefunctions, $\rho_{\mathrm{MZM}}^{(i)}$, depicted in Figs.~\ref{fig:figure2}, \ref{fig:figure4}, and \ref{fig:figure6} we define the spin dependent on-site MZM operators
\begin{align} \label{eq:majorana_ops}
    w_\sigma &= (d_\sigma + d_\sigma^\dagger) \\
    z_\sigma &= i(d_\sigma - d_\sigma^\dagger).
\end{align}
The MZM on the ABS are defined analogously with the corresponding creation and annihilation operators.
From Eq.~\eqref{eq:majorana_ops} we find the MZM wavefunctions through 
\begin{align}
    \rho_{\mathrm{MZM}}^{(w)}&= | \langle \psi_{\mathrm{odd}}^{(0)}|w_\sigma+w_{\overline{\sigma}}|\psi_{\mathrm{even}}^{(0)}\rangle |^2 \\
    \rho_{\mathrm{MZM}}^{(z)}&= | \langle \psi_{\mathrm{odd}}^{(0)}|z_\sigma+z_{\overline{\sigma}}|\psi_{\mathrm{even}}^{(0)}\rangle |^2 
\end{align}
where $|\psi^{(\lambda)}\rangle$ denotes an eigenstate of the many-body Hamiltonian, cf. Eq.~\eqref{eq:H_DA}.
The conductance plot in Fig.~\ref{fig:figure2} e) has been obtained by implementing the rate equations listed in the supplementary material of \cite{Tsintzis2022Creating}. The energies of the Kitaev model \cite{Kitaev2001Unpaired} in Figs.~\ref{fig:figure4}, and \ref{fig:figure6} have been obtained by exact diagonalization and replacing the parameters by their corresponding partners from perturbation theory, cf. Eq.~\eqref{eq:kitaev_replacements}. Lastly, the perturbative analysis has in parts been performed by using Pymablock \cite{Day2024Pymablock}. All codes related to the results presented in the manuscript can be found in the accompanying Zenodo repository \cite{DotABSZenodo}

\section{Symmetries of the charge stability diagram}\label{sec:symmetry}
\begin{figure}[h]
    \centering
    \includegraphics[width=0.5\linewidth]{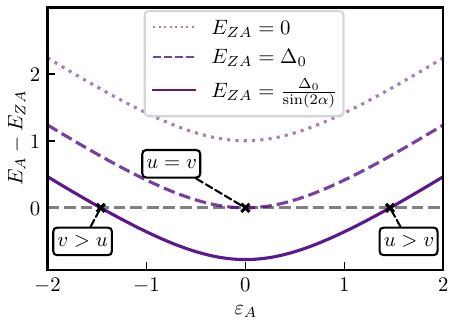}
    \caption{ABS energy depending on the on-site chemical potential of the ABS dot for varying Zeeman energies $E_{ZA}$. Depending on the Zeeman energy, at most two ABS solutions will become available to potentially host sweet spots of the system.}
    \label{fig:figure8}
\end{figure}

The CSD depicted in Fig.~\ref{fig:figure2} shows six distinct regions with alternating groundstate properties.
If the low-energy subspace is described by the Hamiltonian given in Eq.~\eqref{eq:H_DA_eff}, then varying the chemical potentials $\varepsilon_i$ on the two sites changes the charge state of the dots.
The possible charge states on each site depending on the chemical potential are
\begin{align}
    &|\downarrow\uparrow\rangle, &&\varepsilon_D < -U_D-E_{ZD} \\
    &|\downarrow\rangle, &&-U_D-E_{ZD} < \varepsilon_D < E_{ZD}  \\
    &|0\rangle, &&\varepsilon_D > E_{ZD}
\end{align}
on the normal dot, and
\begin{align}
    &u|0\rangle+v|\uparrow \downarrow\rangle, &&\varepsilon_A < -\sqrt{E_{ZA}^2-\Delta_0^2} \label{abs_state_le}\\
    &|\downarrow\rangle, &&-\sqrt{E_{ZA}^2-\Delta_0^2} < \varepsilon_A < \sqrt{E_{ZA}^2-\Delta_0^2}  \\
    &u|0\rangle+v|\uparrow \downarrow\rangle,  &&\varepsilon_A > \sqrt{E_{ZA}^2-\Delta_0^2}, \label{abs_state_he}
\end{align}
on the proximitized dot, where $u>v$ in Eq. \eqref{abs_state_le}, and $v>u$ in Eq. \eqref{abs_state_he}.
The precise choice of the corner in the CSD depends on 1) the choice of the degeneracy on the normal dot, and 2) the choice of the ABS, i.e. $\varepsilon_A=\pm\sqrt{E_{ZA}^2-\Delta_0^2}$.
The choice of the normal dot degeneracy yields different low energy Hamiltonians due to the different spin orientations that are relevant for the transition.
For the right two corners, the normal dot orbital is spin-down, so we have
\begin{align}
H_T &\approx -t_{so} c\dg_{\su} d_{\sd} + t c\dg_{\sd} d_{\sd} + h.c. \nn
&\approx -t_{so} v \gamma_{\sd} d_{\sd} - t u \gamma\dg_{\sd} d_{\sd} + h.c.,
\end{align}
where we have used the Bogoliubov transformation
\begin{align}
 c\dg_{\su} = u \gamma\dg_{\su} + v \gamma_{\sd},\quad c\dg_{\sd} = -u \gamma\dg_{\sd} + v \gamma_{\su}.
\end{align}
So we have 
\begin{align} \label{app:right_effs}
\teff = -tu, \quad \Deff = -t_{so}v
\end{align}
for both bottom-right and top-right corners.
On the other hand, for the left two corners, the dot orbitals are spin-up states, giving 
\begin{align}
H_T &\approx t^* c\dg_{\su} d_{\su} + t_{so} c\dg_{\sd} d_{\su} + h.c. \nn
&\approx t^* v \gamma_{\sd} d_{\su} - t_{so} u \gamma\dg_{\sd} d_{\su}.
\end{align}
Thus 
\begin{align}
\teff = -t_{so}u, \quad \Deff = t^* v.
\end{align}
The choice top- or bottom-corner depends on the choice of the ABS.
In Fig.~\ref{fig:figure8} we have depicted the different options in the parameter regimes relevant for the problem.
Choosing the negative energy ABS ($\varepsilon_A=-\sqrt{E_{ZA}^2-\Delta_0^2}$) yields $v>u$ while it is the opposite ($u>v$) for the positive energy ABS. 
Finally, the availability of a corner to host a sweet spot depends on the relation between $t$ and $t_{so}$.
The main text discusses the behavior at the bottom-right sweet spot when $t>t_{so}$, a condition that needs to be satisfied for the bottom-right corner to be a viable sweet spot. 
The choice of corner is then determined by the ABS, i.e. for the bottom-right corner that is $\varepsilon_A=-\sqrt{E_{ZA}^2-\Delta_0^2}$ which yields $v>u$.
This is necessary to enable $|t_{\mathrm{eff}}|=|\Delta_{\mathrm{eff}}|$, see Eq.~\eqref{app:right_effs}.
Choosing the opposite ABS, i.e. $\varepsilon_A=\sqrt{E_{ZA}^2-\Delta_0^2}$, one switches from the bottom right to the top right corner.
This corner can however not host any sweet spot when $t>t_{\mathrm{so}}$ since
\begin{align}
    u&=\sqrt{\frac{1}{2}\left(1+\sqrt{1-\Delta_0^2/E_{ZA}^2}\right)}\overset{\frac{\Delta_0}{E_{ZA}}\nearrow 1}{\approx}\frac{\sqrt{2}}{2}+\frac{\sqrt{1-\Delta_0^2/E_{ZA}^2}}{2}+\frac{\sqrt{2}}{8}\left(\frac{\Delta_0^2}{E_{ZA}^2}-1\right) \\
    v&=\sqrt{\frac{1}{2}\left(1-\sqrt{1-\Delta_0^2/E_{ZA}^2}\right)}\overset{\frac{\Delta_0}{E_{ZA}}\nearrow 1}{\approx}\frac{\sqrt{2}}{2}-\frac{\sqrt{1-\Delta_0^2/E_{ZA}^2}}{2}+\frac{\sqrt{2}}{8}\left(\frac{\Delta_0^2}{E_{ZA}^2}-1\right)
\end{align}
showing that $u>v$ for $E_{ZA}>\Delta_0$ (see also Fig.~\ref{fig:figure8}).
If however $t_{\mathrm{so}}>t$, the constraint on $u, v$ is inverted and the availability of the two corners flips.
For clarity, we introduce the precise definitions by which we refer to the corners of the CSD in table~\ref{tab:csd_corners}. 
There, we list the participating, i.e. degenerate states, on the normal dot (dot $D$) and the proximitized dot (dot $A$).
Furthermore, we give the conditions relevant for the existence of the sweet spot, e.g. that a bottom right sweet spot becomes viable if both $v>u$ and $t>t_{\mathrm{so}}$ are met.
\begin{table}[h]
    \centering
    \begin{tabular}{c|c|c}
         & \textbf{Left} & \textbf{Right} \\
         & Dot $D$: $|\downarrow\uparrow\rangle$ \; ; $|\downarrow\rangle$ & Dot $D$: $|\downarrow\rangle$ \; ; $|0\rangle$ \\ \hline
         \textbf{Top} & $t>t_{\mathrm{so}}$ & $t<t_{\mathrm{so}}$ \\
          Dot $A$: $u|0\rangle+v|\uparrow \downarrow\rangle$ \; ; $|\downarrow\rangle$ & $u>v$  & $u>v$ \\ \hline 
         \textbf{Bottom} &$t<t_{\mathrm{so}}$ &$t>t_{\mathrm{so}}$ \\
         Dot $A$: $|\downarrow\rangle$ \; ; $u|0\rangle+v|\uparrow\downarrow\rangle$& $u<v$ & $u<v$ \\ \hline
         $H_{DA}^{\mathrm{eff}}$ terms& $t_{\mathrm{eff}}=-t_{\mathrm{so}}u \; ; \Delta_{\mathrm{eff}}=t^\dagger v$ & $t_{\mathrm{eff}}=-tu \; ; \Delta_{\mathrm{eff}}=-t_{\mathrm{so}}v$ \\
    \end{tabular}
    \caption{Definition of the corners visible in the charge stability diagram, Fig.~\ref{fig:figure2}. The given constraints on $u,v$ and $t, t_{\mathrm{so}}$ determine the whether the corresponding corner in the charge stability diagram is a viable sweet spot.}
    \label{tab:csd_corners}
\end{table}

\section{Finite charging energy $U_{\mathrm{ABS}}$ in the Andreev bound state}\label{sec:finite_U}
The main text discusses a configuration without charging energy on the ABS dot. 
Removing this constraint, we obtain the Hamiltonian on the ABS dot as
\begin{align}
    H_A &= (\varepsilon-E_{ZA})c^\dagger_\downarrow c_\downarrow +(\varepsilon+E_{ZA})c^\dagger_\uparrow c_\uparrow + \Delta_0 c^\dagger_\uparrow c^\dagger\downarrow + U_{\mathrm{ABS}} c^\dagger_\uparrow c^\dagger_\downarrow c\downarrow c\uparrow + \mathrm{h.c}
\end{align}
in the electronic basis. 
$U_{\mathrm{ABS}}$ is the charging energy on the ABS and the remaining symbols are defined in Sec.~\ref{sec:model_hamiltonian}. 
In the many-body basis, $\{|0\rangle,|\downarrow \uparrow\rangle,|\downarrow\rangle,|\uparrow\rangle\}$, we can write the Hamiltonian
\begin{align}
    H_A &= \begin{pmatrix}
        0 & \Delta & 0 & 0 \\
        \Delta & 2\varepsilon+U_{\mathrm{ABS}} & 0 & 0 \\
        0 & 0 & \varepsilon-E_{ZA} & 0 \\
        0 & 0 & 0 & \varepsilon+E_{ZA}
        \end{pmatrix}.
\end{align}
We substitute $\xi=2\varepsilon+U_{\mathrm{ABS}}$ and define $E_\xi=\sqrt{\xi^2+\Delta^2}$. 
With these replacements we can write the groundstate of the even and odd parity subspaces as
\begin{align}
    &E_{\mathrm{GS}}^{(even)}=\xi-\sqrt{\xi^2+\Delta^2}; \quad &&|\mathrm{GS}_{odd}\rangle=\sqrt{\frac{E_\xi+\xi}{2E_\xi}}|0\rangle + \sqrt{\frac{E_\xi-\xi}{2E_\xi}}|\uparrow \downarrow\rangle \\
    &E_{\mathrm{GS}}^{(odd)}=\varepsilon-E_{ZA}; \quad && |\mathrm{GS}_{even}\rangle=|\downarrow\rangle
\end{align}
To induce MZMs on the dots, the two groundstates need to be degenerate. 
They are connected through a quasiparticle excitation $|\mathrm{GS}_{even}\rangle=(uc^\dagger_\downarrow-vc^\dagger_\uparrow)|\mathrm{GS}_{odd}\rangle$ with $u, v$ to be determined.
We obtain the condition
\begin{align}
    u\sqrt{\frac{E_\xi+\xi}{2E_\xi}}+v\sqrt{\frac{E_\xi-\xi}{2E_\xi}}=1
\end{align}
which is solved by $u(U_{\mathrm{ABS}})=\sqrt{\frac{E_\xi+\xi}{2E_\xi}}, v(U_{\mathrm{ABS}})=\sqrt{\frac{E_\xi-\xi}{2E_xi}}$ since $u(U_{\mathrm{ABS}})^2+v(U_{\mathrm{ABS}})^2=1$. 
The degeneracy condition requires
\begin{align}
    E_{\mathrm{GS}}^{(even)}=E_{\mathrm{GS}}^{(odd)}
\end{align}
leading to $\xi$ being constrained to
\begin{align}
    \xi^2=\left(\frac{U_{\mathrm{ABS}}}{2}+E_{ZA}\right)^2-\Delta^2.
\end{align}
To satisfy $v(U_{\mathrm{ABS}})>u(U_{\mathrm{ABS}})$ (see App.~\ref{sec:symmetry}) we choose the negative root solution for $\xi$.
Gathering all findings into $u(U_{\mathrm{ABS}})$ and $v(U_{\mathrm{ABS}})$ we can write
\begin{align}
    u(U_{\mathrm{ABS}})&=\sqrt{\frac{E_\xi+\xi}{2E_\xi}}\overset{U_{\mathrm{ABS}}\rightarrow 0}{\approx}u-\frac{u}{4}\frac{\Delta^2}{\xi_0E_{ZA}(E_{ZA}-\xi_0)}U_{\mathrm{ABS}} \\
    v(U_{\mathrm{ABS}})&=\sqrt{\frac{E_\xi-\xi}{2E_\xi}}\overset{U_{\mathrm{ABS}}\rightarrow 0 }{\approx}v+\frac{v}{4}\frac{\Delta^2}{\xi_0E_{ZA}(E_{ZA}-\xi_0)}U_{\mathrm{ABS}}
\end{align}
where we used $\xi_0=\sqrt{E_{ZA}^2-\Delta^2}$, and $u, v$ as defined in Sec.~\ref{sec:model_hamiltonian}.
We recognize that $u$ decreases while $v$ increases.
It is therefore to be expected that the Zeeman energy at which the sweet spot is observed reduces.
Indeed we find $E_{ZA}^*$ for $\theta=0$ at the sweet spot as
\begin{align} \label{eq:ez_ss_u_abs}
    E_{ZA}^*=\frac{\Delta_0}{\sin(2\alpha)}-\frac{U_{\mathrm{ABS}}}{2}.
\end{align}
Since two corners of the charge stability diagram are roughly separated by $\simeq 2E_{\mathrm{ZA}}+U_{\mathrm{ABS}}$, a finite $U_{\mathrm{ABS}}$ can serve to help make the sweet spot more resilient towards single parameter perturbations.
Furthermore, Eq.~\eqref{eq:ez_ss_u_abs} shows that large enough $U_{\mathrm{ABS}}$ can push the sweet spot Zeeman energy below $\Delta_0$.
Hence, sweet spots can emerge even if $E_{\mathrm{ZA}}<\Delta_0$ as a result of the separation of the ABS states in $\varepsilon_{A}$ and Eq.~\eqref{eq:ez_ss_u_abs}.
This final property might be particularly useful when the platform inherent g-factor might be limited through other constraints.

\section{Controlling the effective pairing through $t_0$} \label{sec:tuning_t0}
\begin{figure}[b]
    \centering
    \includegraphics[width=\textwidth]{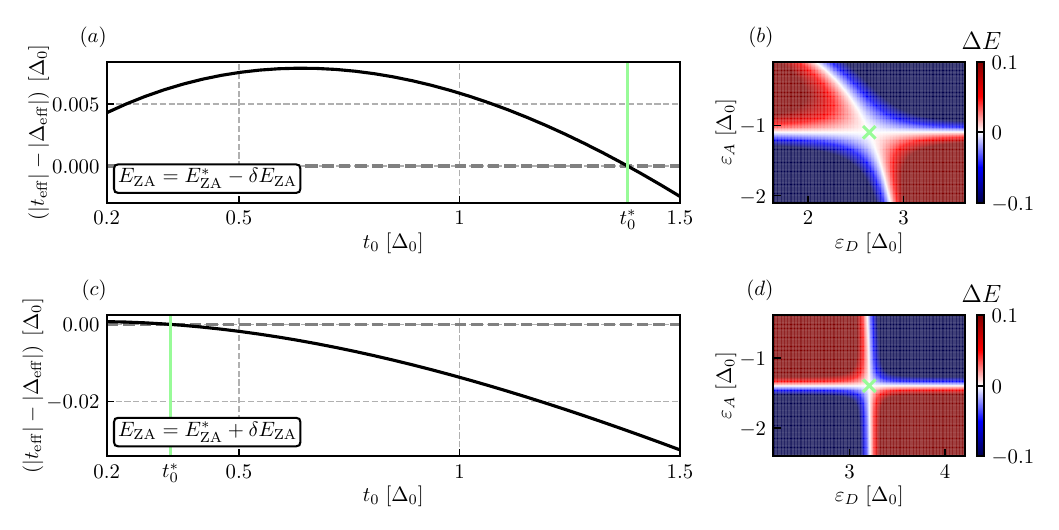}
    \caption{sweet spot recovery through adjustment of $t_0$. a) and b) show how, when starting with $t_0=0.2$ but $E_{\mathrm{ZA}}=E_{\mathrm{ZA}}^*-\delta E_{\mathrm{ZA}}$, a sweet spot can be recovered by increasing $t_0$. Particularly b) demonstrates how the corresponding $t_0>\Delta_0$, strongly coupling the two dots. We lastly want to highlight that, despite the slope of $|t_{\mathrm{eff}}|-|\Delta_{\mathrm{eff}}|$ being negative towards smaller $t_0$ in a), it is impossible to recover a sweet spot by decreasing the hopping further. c) and d) demonstrates sweet spot recovery when $t_0=1.5$ initially and $E_{\mathrm{ZA}}=E_{\mathrm{ZA}}^*+\delta E_{\mathrm{ZA}}$.}
    \label{fig:figure9}
\end{figure}

In Sec.~\ref{sec:proximity_dot} we used a phenomenological pairing parameter $\Delta_D$ on the initially normal dot to demonstrate how a change on the pairing on the dot can recover a sweet spot.
This section appends to Sec.~\ref{sec:proximity_dot}, demonstrating explicitly on the many-body system how the induced pairing from the ABS is controlled through the bare hopping $t_0$. 
In particular,, when the ABS dot is tuned slightly off the sweet spot in $E_{\mathrm{ZA}}$, we can adjust $t_0$ to recover a sweet spot.
Fig.~\ref{fig:figure9} demonstrates sweet spot recovery along two examples: a) and b) show how large $t_0$ recover a sweet spot when starting with an initially small $t_0 (=0.2)$ and $E_{\mathrm{ZA}}=E_{\mathrm{ZA}}^*-\delta E_{\mathrm{ZA}}$; c) and d) demonstrate recovery for small $t_0$ when beginning with $t_0=1.5$, i.e. strong coupling of the dots, at $E_{\mathrm{ZA}}=E_{\mathrm{ZA}}^*+\delta E_{\mathrm{ZA}}$.

\section{Inhomogeneity in the dot-ABS array} \label{app:inhom}
In this section, we consider the effect of Hamiltonian parameter inhomogeneity in a dot-ABS array.
This captures the realistic situation of an experimental device.
To demonstrate the main physical effect, we focus on the three-site Kitaev chain with four different scenarios: DAD and ADA with inhomogeneous spin-orbit interaction, and DAD and ADA with inhomogeneous $g$ factor. 
Here we choose the level of inhomogeneity to be $10\%$ to generate the results in Figs.~\ref{fig:app_inhom_so}and~~\ref{fig:app_inhom_z} and we emphasize that our results are robust even for larger values.
We find that the physical findings and main conclusions presented in the main text are still valid, e.g., the presence of long-range coupling between Majoranas with negligible wavefunction overlap, and energy spectra against chemical potential detuning.
In our simulation here, we need to first figure out the sweet spots in each two-site DA pair by varying the tunnel strength $t_0$.
After that, the sweet spot of a three-site one is obtained by putting them together.
It is likely that the middle site (either dot or ABS) may reach different values of chemical potential for the left and right pairs respectively, and we choose to take the average of them.
In addition, to capture the proximity effect from the continuum states, we add a pairing term $\Delta_{D, induced} \approx t^2_0 \delta \Delta$ on normal quantum dots, with $\delta \Delta= 0.5 \Delta_0$.
We note that the induced gap is proportional to $t^2_0$ owing to second-order process of local Andreev reflection, and that $\delta \Delta$ is a phenomonological parameter which is proportional to the superconductor density of states. 
We have checked that our simulation results do not depend on the precise value of $\delta \Delta$.
To summarize, we have shown that even in the presence of parameter inhomogeneity, the sweet spot of an extended Kitaev chain can still be found by varying the tunnel strength between dot and ABS.
Furthermore, the main findings presented in the main text are still valid.

\subsection{Inhomogeneous spin-orbit mixing}
\begin{figure}
    \centering
    \includegraphics[width=\linewidth]{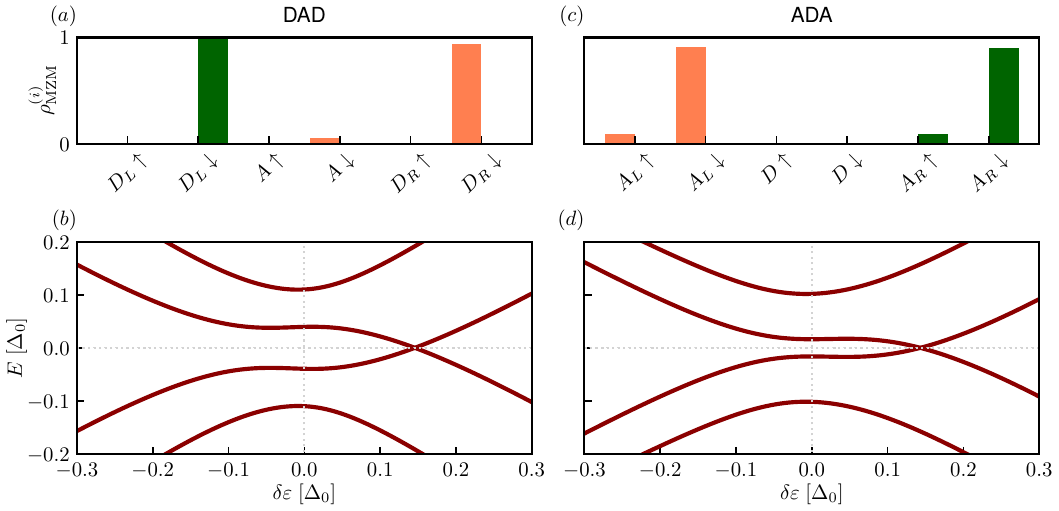}
    \caption{Inhomogenous spin-orbit mixing between the two pairs. We allow for a deviation of $10\%$ between the different $\alpha_i$ ($\alpha_L=0.3, \alpha_R=0.27$). We see that the ADA set-up reproduces the behavior suggested in the main text (d)) despite the inhomogeneity and yields well separated Majoranas (c)). This is explained by the better protection against next-nearest neighbor hopping from the larger g-factor in the central, normal dot. The DAD set-up however is more sensitive to changes of the chemical potential of the middle dot. The smaller g-factor in the ABS dot generally leads to poorer protection against next-nearest neighbor hopping. Yet, we obtain still well separated Majoranas despite the sweet spot only being meta stable against global changes of the chemical potential.}
    \label{fig:app_inhom_so}
\end{figure}

We first consider inhomogeneous spin-orbit mixing in both ADA and DAD set-ups.
We assume that $\alpha_L$ and $\alpha_R$, i.e. the spin-orbit mixing angles of the two pairs.
The results of this analysis are depicted in Fig. \ref{fig:app_inhom_so}.
We recognize that, for both set-ups, we can well recover the spectral behavior discussed in Sec. \ref{sec:scaling_up}.
Furthermore, we see that the Majorana wavefunctions shown in Fig. \ref{fig:app_inhom_so} a) and c) are still well separated.

\subsection{Inhomogeneous $g$-factor}
\begin{figure}
    \centering
    \includegraphics[width=\linewidth]{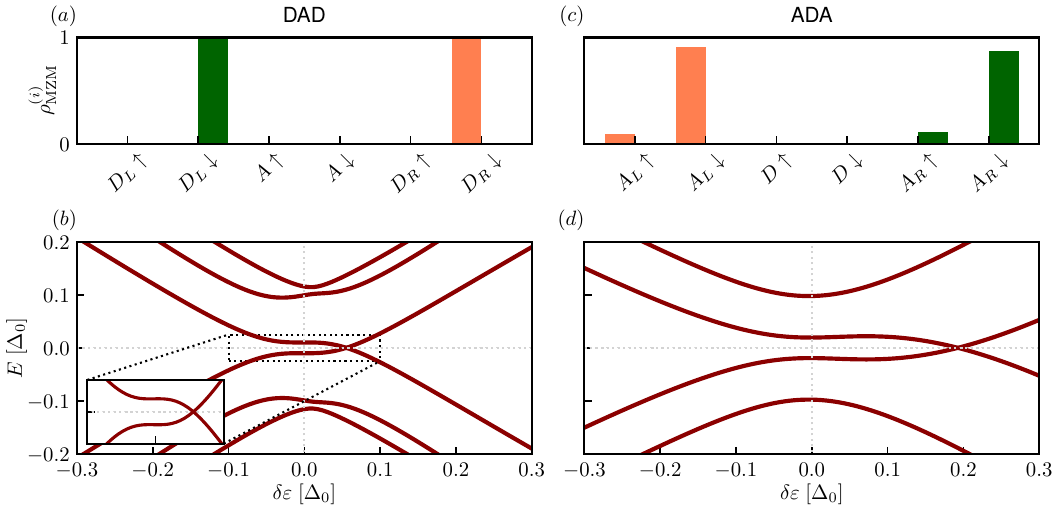}
    \caption{Inhomogenous g-factors between the dots. For the DAD set-up, a) and b), we let the inhomogeneity be as large as $10\%$ ($g_L=2, g_R=1.8$). We find that, despite the stark difference, the spectrum reproduces the findings of the main text well and the Majoranas yielded by the system are well separated from each other. This resillience is due to the larger level separation of the levels on the normal dots. For the ADA set-up we allow for an inhomogeneity of $10\%$ ($g_L=1, g_R=0.9$). In both cases we recover spectral lines akin to those demonstrated in the main text despite the presence of inhomogeneities. Consequently, also the Majorana wavefunctions remian well separated between the dots.}
    \label{fig:app_inhom_z}
\end{figure}

Lastly, we consider inhonmogenous g-factors between the dots. 
The different g-factors of the outer dots between the two set-ups makes them differently susceptible to inhomogeneities of the g-factor. 
We choose $g_L=2, g_R=1.8$ for the DAD and $g_L=1, g_R=0.9$ for the ADA set-up.
This choice yields reasonably different sweet spots of the two pairs of 2-site chains that still allow to be connected by barrier tuning of the second pair, i.e. varying $t_0$.
The results of this analysis are depicted in Fig. \ref{fig:app_inhom_z}
For both situations we find that the behavior of the spectrum suggested in the main text can still be reasonably well reproduced and that the Majoranas that the systems yield are still well separated. 

\end{document}